\documentclass{article}
\usepackage[T1]{fontenc}
\usepackage[utf8]{inputenc}
\usepackage[english]{babel}
\usepackage{graphicx}
\usepackage{todonotes}
\usepackage{hyperref}
\usepackage{float}
\usepackage{amsmath}
\usepackage{ stmaryrd }
\usepackage{amssymb}
\usepackage{tikz}
\usepackage{xcolor}
\usepackage{leftindex}
\usepackage{enumitem}
\usepackage{mathtools}
\usepackage{slashed}
\usepackage{afterpage}
\usepackage{mathrsfs}
\usepackage{blindtext}
\usepackage{titlesec}
\usepackage{biblatex}
\usepackage{textpos}
\usetikzlibrary{positioning}
\usetikzlibrary{calc}
\usepackage{soul}
\usepackage{tikz}
\usepackage{tikz-feynman}
\tikzfeynmanset{compat=1.1.0}
\DeclareMathOperator{\loc}{loc}

\DeclareMathOperator{\sgn}{sgn}
\DeclareMathOperator{\supp}{supp}
\DeclareMathOperator{\sd}{sd}
\DeclareMathOperator{\ret}{ret}

\DeclareMathOperator{\op}{op}
\DeclareMathOperator{\singsupp}{sing \,\, supp}
\DeclareMathOperator{\WF}{WF}
\usepackage{authblk}

\title{Causal perturbation theory and scattering amplitudes}
\author[1]{H.A.C. Grande\thanks{henriqueay@usp.br}}
\author[1]{J.C.A Barata}
\affil[1]{Instituto de Física da Universidade de São Paulo, \protect\\ 
Rua do Matão 1371, São Paulo, Brazil}
\date{September 2025}                     %% if you don't need date to appear
\begin{document}

\maketitle

\begin{abstract}

Motivated by the limited interaction between the mathematical physics community and theoretical physicists—particularly in high-energy theory—we present a computation that is typically the first example in QFT courses but, to our knowledge, does not appear in the literature: the scattering amplitudes of the $\lambda\phi^4$ model in four-dimensional space-time, derived within the framework of causal perturbation theory. Our aim is to introduce this mathematically rigorous formalism in the most accessible way possible. To that end, we emphasize general aspects of the theory while deliberately avoiding overly sophisticated mathematics. Finally, we briefly discuss how the divergent integrals encountered in quantum field theory can be reinterpreted as issues concerning the domain of distributions.

\end{abstract}

\tableofcontents

\section{Introduction}

Quantum field theory (QFT) is the most accurately tested physical theory known, yet many of its aspects remain poorly understood. One of the major challenges in QFT is the scarcity of exactly solvable models. Because of this, perturbative quantum field theory has become the main working framework. The approach, developed primarily by Feynman, Schwinger, Tomonaga, and Dyson in the 1940s, based on the earlier theory of electrodynamics by Heisenberg and Pauli in the late 1920s, is the one adopted in most QFT courses and is the most widely used in research articles. Nevertheless, it suffers from a lack of mathematical rigor. The construction relies on the so-called ``interaction picture'', whose non-existence was rigorously proved by Haag in 1955 \cite{Haag:212242}. Perhaps even more astonishingly, its remarkably precise predictions are obtained through manipulations of ``infinities'' in a rather ad hoc manner.\\

The Feynman approach is the most famous, but it is not the only one. In the early 1940s, Heisenberg proposed a simple theory of the $S$-matrix \cite{Heisenberg1943kf,Heisenberg1943dc,Heisenberg1944dg}, aiming to find a non-perturbative solution to the scattering problem. Although Heisenberg himself could not solve the problem, his series of papers opened the door for Stückelberg to develop a more consistent theory of perturbative scattering in the following decade. His work was largely overlooked by the broader physics community at the time, except by the Soviet physicist Nikolay Bogoliubov. Bogoliubov further developed the theory until it was mature enough for H. Epstein and V. Glaser to finally formulate a mathematically rigorous ``recipe'' \cite{epstein1973role} that reproduced the same results as those obtained in the celebrated Feynman formalism. Besides its mathematical clarity, this construction avoided divergent integrals and therefore did not require any artificial ``cut-off.''\\

\begin{figure}[H]
    \centering
    \includegraphics[width=0.5\linewidth]{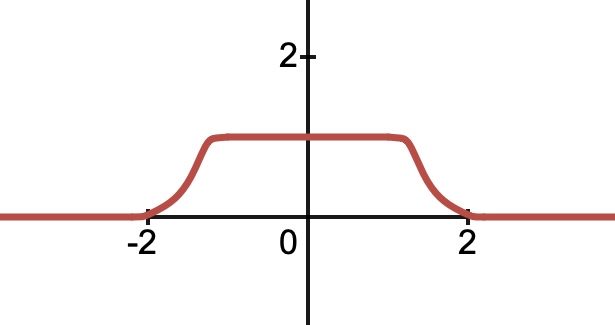}
    \caption{Example of a test function. We have $g(x)=1$ in the ``interaction'' region.}
    \label{fig:test-function}
\end{figure}

To maintain consistency, however, the theory must be constructed locally. In practice, this means that the interactions $\mathcal{L}(x)$ must be multiplied by a smooth, compactly supported test function $g(x)$ (see Figure \ref{fig:test-function}), which equals one in the region where the interaction is ``turned on'' and vanishes elsewhere. The subtlety of this construction lies in taking the so-called adiabatic limit $g(x)=1 \ \forall x$ \cite{duch2017masslessfieldsadiabaticlimit} (additional historical remarks can be found in \cite{perturbativecausality,Fraser2024-kp}).\\

More recently, the causal approach has been extended to curved spacetimes and further refined. Its current form incorporates deformation quantization, a state-independent framework, and a close connection with algebraic quantum field theory \cite{AdvancesInAlgebraicQuantumFieldTheory}.\\

There is abundant material for those seeking a deeper understanding of the aforementioned topics (see, e.g., \cite{book,c37b9fad8f104637ad2b22a76f1e1e1e}). However, except for an experienced reader, these references are generally difficult to follow. With this in mind, we have written this article with the goal of deriving the Feynman rules for the scattering of neutral scalar fields in the simplest possible language. We assume that the reader has some familiarity with the theory of generalized functions \cite{hörmander1983analysis}, although deep knowledge is not required. From the physical side, we assume a basic understanding of quantum field theory in the Fock space setting. The goal of the present article is \textbf{not} to provide a complete treatment of the subject, but rather to pique the reader's curiosity and encourage them to seek further references. This article is result of the master's project of the author, were the computations have been done with greater detail \cite{grande2025}. \\

The article is divided as follows: in section two we introduce the notation and discuss briefly the concept of a distribution. The third section is dedicate to classical fields and it is followed by the quantization of it. Finally,in the last chapter it is shown how to construct the $T-$ product as well as compute the scattering matrix.

%%%%%%%%%%%%%%%%%%%%%%%%%%%%%%%%%%%%%%%%%%%%%%%%%%%%%%%%%%%%%%%%%%%%%
\section{Propagators and Conventions}

This section provides a concise review of the underlying structure of the theory. Our main goal here is to fix the notation and conventions that will be used throughout the paper. For this reason, we also present the propagators in advance. Their motivation will be briefly discussed here, while their full derivation will appear later in the appropriate context. 

\subsection{Minkowski Space-Time}

Unless explicitly stated otherwise, the physical background of this work is the $d$-dimensional Minkowski space-time, denoted by $\mathbb{M}$. A vector $x\in\mathbb{M}$ is written as $x:=(x^0,\vec{x})$. We adopt the metric convention
\[
g \equiv \eta = \mathrm{diag}(1,-1,-1,\dots,-1).
\]
Accordingly, the inner product is written as
\[
px \equiv p_\mu x^\mu = p^0x^0 - \vec{p}\cdot\vec{x}.
\]

We define the \textbf{forward and backward light cones} as
\begin{gather}
    V_+ := \{x\in\mathbb{M}\,|\,x^2>0,\; x^0>0\}, 
    \quad 
    V_- := \{x\in\mathbb{M}\,|\,x^2>0,\; x^0<0\}.
\end{gather}

Their closures are denoted by $\overline{V}_\pm$. The \textbf{thin diagonal} is defined as
\begin{gather}
    \Delta_n := \{(x_1,\dots,x_n)\in\mathbb{M}^n \,|\, x_1=x_2=\cdots=x_n\}.
\end{gather}

We work in natural units,
\begin{gather}
    \hbar = c = 1.
\end{gather}

Nevertheless, we introduce a parameter $\hbar$ later on. This parameter should not be interpreted as the physical Planck constant, but rather as a bookkeeping device fixed to $\hbar \in \{0,1\}$. The value $\hbar=1$ corresponds to the ``quantum theory'', while $\hbar=0$ corresponds to the ``classical theory''\footnote{In the text we often write $\hbar\to 0$ or $\hbar\to 1$ instead of $\hbar=0,1$. The idea is to speak of ``taking the classical/quantum limit''. Since no topology has been specified, this limit should be understood as simply replacing $\hbar$ by $0$ or $1$.}. 

\subsection{Formal Power Series and Notation}

As is standard in perturbative approaches, most physically relevant quantities are expressed as formal power series in some constant $\lambda$. Let $\mathcal{V}$ be a complex vector space and $\lambda\in\mathbb{R}$. The \textbf{set of formal power series in $\lambda$ with coefficients in $\mathcal{V}$} is defined as
\begin{gather}
    \mathcal{V}\llbracket \lambda \rrbracket 
    := \left\{ V \equiv \sum_{n=0}^\infty \lambda^n V_n \equiv (V_n)_{n\in\mathbb{N}} \,\middle|\, V_n\in\mathcal{V} \right\}.
\end{gather}

In general, in quantum field theory, the convergence of such series is not controlled. We also adopt the following notation for the d’Alembertian:
\begin{gather}
    \partial^2 \equiv \square \equiv \partial_\mu \partial^\mu 
    = \partial_t^2 - (\vec{\nabla})^2.
\end{gather}

\subsection{Distributions}

The mathematical framework employed in this paper is the theory of distributions (generalized functions). Classical references include \cite{reed_simon_Vol1} and \cite{hörmander1983analysis}. \\

Let $\Omega\subseteq\mathbb{M}$. The set of \textbf{smooth functions} $f:\Omega\to\mathbb{R}$, i.e., infinitely differentiable functions, is denoted by
\begin{gather}
    \mathcal{C} := C^\infty(\mathbb{M},\mathbb{R}) \equiv C^\infty(\mathbb{R}^d,\mathbb{R}).
\end{gather}

The \textbf{support} of a function is defined as
\begin{gather}
    \supp(f) := \overline{\{x\in\Omega \,|\, f(x)\neq 0\}}.
\end{gather}

The set of smooth functions with compact support is denoted $\mathcal{D}(\Omega)$. An element $f\in \mathcal{D}(\Omega)$ is called a \textbf{test function}.\\

A \textbf{distribution} is a continuous linear functional $t:\mathcal{D}(\Omega)\to\mathbb{R}$. The set of all such distributions is denoted by $\mathcal{D}'(\Omega)$. For $t\in\mathcal{D}'(\Omega)$ and $g\in\mathcal{D}(\Omega)$ we write
\begin{gather}
    t(g) \equiv \langle t,g\rangle \equiv \langle t(x),g(x)\rangle_x
    \equiv \int_\Omega dx\, t(x)g(x).
\end{gather}

For several variables, we use the shorthand notation
\begin{gather}
    dx_1\cdots dx_n \equiv dX_n.
\end{gather}

If the domain of integration is not specified, it is understood to be the whole space. The support of a distribution $t\in\mathcal{D}'(\Omega)$ is the smallest closed set $K\subseteq\Omega$ such that $t|_{\mathcal{D}(\Omega\setminus K)}=0$, and is also denoted by $\supp(t)$. By abuse of notation we may write $t(x)=0$ to mean $t|_{\mathcal{D}(\Omega)}=0$.\\

\textit{Remark.} Integral notation in this context is symbolic and should not always be interpreted as an actual integration over a domain.

\section{Examples of distributions and propagators}

Some of the most commonly used distributions in physics are $\delta(x)\in\mathcal{D}'(\mathbb{R})$ and $\theta(x)\in\mathcal{D}(\mathbb{R})$, defined as follows:

\begin{gather}
\langle\delta(x),f(x)\rangle \equiv \int dx\, \delta(x)f(x) = f(0), \nonumber\\
\langle\theta(x),f(x)\rangle \equiv \int dx\, \theta(x)f(x) = \int_0^\infty dx\, f(x).
\end{gather}

The Heaviside distribution $\theta(x)$ can also be expressed in terms of an ordinary function:

\begin{gather}
\theta(x) :=
\begin{cases}
1 & x \geq 0,\\
0 & x < 0.
\end{cases}
\label{Heavyside Theta}
\end{gather}

An important result states that if $t \in \mathcal{D}'(\mathbb{R}^d)$ with $\supp(t)\subseteq {0_d}$, where $0_d=(0,0,\dots,0)$ denotes the zero vector in $d$ dimensions, then

\begin{gather}
t(x) = \sum_{a} C_a \partial^{a} \delta_d(x),
\end{gather}

where $\delta_d$ is the Dirac delta distribution in $d$ dimensions, $C_a \in \mathbb{C}$, and the sum is finite \cite[p.~46, Theorem 2.3.4]{hörmander1983analysis}.

The next family of examples, known as \textbf{propagators}, arises from their physical interpretation. The propagators presented here correspond to the free theory of a neutral scalar field $\phi$. A more detailed discussion will be provided later.

The \textbf{retarded propagator} $\Delta^{\ret}(x)\in\mathcal{D}'(\mathbb{R}^d)$ is defined by:

\begin{gather}
    \Delta^{\ret}(x):=\frac{1}{(2\pi)^d}\int d^dp\, \frac{e^{-ipx}}{p^2-m^2+ip^00},
\end{gather}

where $px=p_\mu x^\mu$ and $ip^00\equiv \lim_{\epsilon\rightarrow0^+} ip^0\epsilon$. The retarded product appears naturally when one tries to find a perturbative solution to an interacting field.\\

The \textbf{Jordan-Pauli function} or \textbf{commutation function} is defined by:

\begin{gather}
    \Delta(x):=\Delta^{\ret}(x)-\Delta^{\ret}(-x)=\frac{-i}{(2\pi)^{d-1}}\int d^dp\, \sgn(p^0)\delta(p^2-m^2)e^{-ipx}.
\end{gather}

This propagator is mostly defined for convenience in practical calculations.\\

The \textbf{Wightman two-point function} also known as the positive part of $i\Delta$ is defined as the propagator of the scalar field $\langle\Omega|\phi(x)\phi(y)\Omega\rangle$. After some work, we can write it as follows:

\begin{gather}
    \Delta^+(x):=\frac{1}{(2\pi)^{d-1}}\int d^dp\, \theta(p^0)\delta(p^2-m^2)e^{-ipx}.
\end{gather}

The Wightman two-point function is important for quantizing the theory. \\

The last propagator that we define is the \textbf{Feynman propagator}:

\begin{gather}
    \Delta^F(x):=\theta(x^0)\Delta^+(x)+\theta(-x^0)\Delta^+(-x)=\frac{i}{(2\pi)^d}\int d^dp\frac{e^{ipx}}{p^2-m^2+i0}.
\end{gather}

The Feynman propagator is important in the construction of the so-called $S-$ matrix.\\

\subsection{Technical Remarks}

Although distributions are often manipulated as if they were ordinary functions, they are not. We recall here some technical tools that will be used later. 

\subsubsection{Derivatives of Distributions}

The derivative of a distribution $t$ is defined by
\begin{gather}
    \langle \partial^a t, g \rangle := (-1)^{|a|}\langle t, \partial^a g\rangle.
\end{gather}

One possible way to explain the formula above is to imagine that the inner product with a test function is indeed an integral and not just notation. If that were the case, we could calculate the derivative of a distribution $t\in\mathcal{D}'(\mathbb{R})$ acting on $g\in\mathcal{D}(\mathbb{R})$ using integration by parts:

\begin{align}
    \int dx\, (\frac{d}{dx}t(x))g(x)\text{ ``}&=\text{'' }t(x)g(x)\bigg\rvert_{-\infty}^\infty-\int dx\, t(x)(\frac{d}{dx}g(x))\nonumber\\
    &=-\int dx\, t(x)(\frac{d}{dx}g(x)).
\end{align}

In the last step, we have used that $g$ is compactly supported and therefore $\lim_{x\rightarrow\pm\infty}g(x)=0$. We reinforce that the integral is only symbolic. One could also use the translation group acting on the distribution to define the derivative of it \cite{Barata39}.

\subsubsection{Fourier Transform}

The Fourier transform of $u\in\mathcal{D}'(\mathbb{M}^n)$, denoted by $\mathcal{F}(u)(k)\equiv \tilde{u}(k)$, is defined analogously:
\begin{gather}
    \langle \mathcal{F}(u)(k),f(k)\rangle_k 
    = \langle u(x),\mathcal{F}(f)(x)\rangle,
    \label{Fourier definition}
\end{gather}
with
\begin{gather}
    \mathcal{F}(f)(x) = \frac{1}{(2\pi)^{\frac{dn}{2}}}
    \int dk\, e^{ikx}f(k).
    \label{fourier transformation}
\end{gather}

\textit{Remark.} The Fourier transform is rigorously defined on the Schwartz space $\mathcal{J}(\mathbb{M}^n)$ of rapidly decaying functions; see the Appendix and \cite{Barata39}. \\

\textit{Remark 2.} Equation \eqref{Fourier definition} can be unfolded as
\begin{align}
    \langle\mathcal{F}(u)(k),f(k)\rangle_k
    &= \int dk \left( \int \frac{dx}{(2\pi)^{\frac{dn}{2}}}
    e^{ikx}u(x)\right) f(k) \nonumber\\
    &= \int dx\, u(x)\left( \int \frac{dk}{(2\pi)^{\frac{dn}{2}}}
    e^{ikx}f(k)\right)
    = \langle u(x),\mathcal{F}(f)(x)\rangle.
\end{align}

The inverse Fourier transform is defined as
\begin{gather}
    \mathcal{F}^{-1}(f)(k) :=
    \int \frac{dx}{(2\pi)^{dn/2}}e^{-ikx}f(x) = \tilde{f}(-k).
\end{gather}

\subsubsection{Multiplication of Distributions: The Wave-Front Set}

In this section, we illustrate explicitly that distributions cannot be treated as ordinary functions. To motivate the discussion, let us consider the following example. Suppose we attempt to compute naively the product $\delta(x)\theta(x)$ applied to a test function $g(x)\in\mathcal{D}(\mathbb{R})$. Using the relation $\theta'(x)=\delta(x)$ introduced above, together with $\theta^2(x)=\theta(x)$ (see \ref{Heavyside Theta}), and applying the Leibniz rule for derivatives, we obtain

\begin{align}
\langle\frac{d}{dx}\theta^2(x),g(x)\rangle&=2\langle\delta(x)\theta(x),g(x)\rangle\overset{!}{=}\langle\frac{d}{dx}\theta(x),g(x)\rangle=\langle\delta,g\rangle\nonumber\\
&\Rightarrow \langle\delta(x)\theta(x),g(x)\rangle= \frac{1}{2}\langle\delta,g\rangle=\frac{1}{2}g(0).
\end{align}

At first sight this appears consistent. However, repeating the same reasoning with $\theta^n(x)$, for any $n>2$, one would obtain

\begin{gather}
\langle\delta(x)\theta(x),g(x)\rangle=\frac{1}{n}\langle \delta,g\rangle=\frac{1}{n}g(0),
\end{gather}

which is clearly contradictory. This simple example highlights the fact that products of distributions must be handled with care and cannot be defined naively. To establish a general framework for the multiplication of distributions, one must rely on the concept of the \emph{wave-front set}. Our presentation will follow mainly \cite{Brouder_2014} and \cite{Barata39}. For a comprehensive treatment of the subject, see \cite{hörmander1983analysis}, Chapters 7 and 8.

Let $u\in\mathcal{D}'(\mathbb{R}^d)$. The \textbf{regular support} of $u$ is the set of points where $u$ coincides with a smooth function. Its complement is the \textbf{singular support}, denoted $\singsupp(u)$. \\

The \textbf{wave-front set} of $u$ is defined as \cite[pg.~474]{book}
\begin{align}
    \WF(u) := \{ (x,k)\in\mathbb{R}^d\times(\mathbb{R}^d\setminus\{0\}) 
    \,|\, x\in\singsupp(u),\;
    \widetilde{uf}\\ \text{ does not decay rapidly in direction }k
    \ \forall f\neq0 \}.
\end{align}

Here, ``rapid decay'' means faster than any polynomial. The wave-front set is a key tool because of the following result.\\

Let $u,v\in\mathcal{D}'(\mathbb{R}^d)$. If
\begin{align}
    (x,0) \notin \WF(u)\oplus\WF(v)
    := \{(x,k_1+k_2)\,|\,(x,k_1)\in\WF(u),\ (x,k_2)\in\WF(v)\},
\end{align}
then the product $uv$ exists and satisfies the Leibniz rule. This result is known as the \emph{Hörmander criterion} (see Theorem 8.2.10, pg.~267 of \cite{hörmander1983analysis}; also \cite{Brouder_2014,book}).\\

\textit{Remark.} In some cases, two distributions may not satisfy the criterion but their product still exists and obeys the Leibniz rule, though this is exceptional. Alternative definitions of products of distributions exist \cite{Different_products}, but they are rarely used in this context.

%%%%%%%%%%%%%%%%%%%%%%%%%%%%%%%%%%%%%%%%%%%%%%%%%%%%%%%%%%%%%%%%%%%%%
\section{Classical fields}

\subsection{Introduction}

In this section we present the formalism that will be adopted throughout the remainder of the work. In the standard approach, fields in Fock space are represented by operators acting on states, yielding distributions as expectation values. Our goal, however, is to construct a framework that does not depend on a particular choice of state. For this purpose, we introduce the notion of a field as a \textit{functional}, namely, an object that takes a function as input and returns a number (real for the scalar field, and complex in the general case).  

The essential difference between this formulation and the usual operator formalism is that, here, the field is not itself an operator and does not necessarily satisfy a ``field equation'' a priori. Instead, the aim is to establish a consistent and self-contained construction. The presentation follows closely the first chapter of \cite{book}.

\subsection{Basic structure}

We define fields as functionals on the configuration space with values in $\mathbb{R}$ or $\mathbb{C}$. Unless stated otherwise, the \textbf{configuration space} is the set of smooth functions on $d$-dimensional Minkowski spacetime,  
\[
C^\infty(\mathbb{M}) \equiv \mathcal{C}.
\]

A \textbf{scalar field} is given by
\begin{align}
    \phi(x):\begin{cases}
        \mathcal{C}\rightarrow\mathbb{R}\\ h\mapsto h(x)
    \end{cases}.
\end{align}

Derivatives of fields are defined in an analogous manner:
\begin{align}
    \partial^a\phi(x):\begin{cases}
        \mathcal{C}\rightarrow\mathbb{R}\\ h\mapsto \partial^ah(x).
    \end{cases}
\end{align}

Here $a$ denotes a multi-index. The \textbf{set of polynomials} generated by $\partial^a\phi$ is denoted by $\mathcal{P}$. 

\subsubsection{The space of fields $\mathcal{F}$}

The \textbf{space of fields} is defined as the set of functionals $F \equiv F(\phi)$ of the form
\begin{align}
    F=f_0+\sum_{n=1}^N\int dx_1\cdots dx_n\, f_n(x_1,\cdots ,x_n)\phi(x_1)\cdots \phi(x_n),
    \label{Standard Field}
\end{align}
with $N<\infty$, where $f_n(x_1,\cdots ,x_n)$ is a $\mathbb{C}$-valued distribution ($f_n \in \mathcal{D}'(\mathbb{M}^d,\mathbb{C})$) satisfying:
\begin{enumerate}
    \item[(i)] symmetry under permutations: $f_n(x_{\pi(1)},\cdots ,x_{\pi(n)})=f_n(x_1,\cdots ,x_n)$ for all $\pi\in S^n$;\\
    \item[(ii)] the wave front set condition:
    \[WF(f_n)\subseteq\left\{(x_1,\cdots ,x_n,k_1,\cdots ,k_n)\,\middle|\, (k_1,\cdots ,k_n)\notin \overline{V}_+^{\times n}\cup \overline{V}_-^{\times n}\right\},\]
    \label{definef}
\end{enumerate}
where $V_\pm$ denote the forward and backward light cones.  

\begin{figure}[H] % use [h] ou [H] conforme sua preferência
    \centering
    \begin{tikzpicture}[scale=0.5, line cap=round, line join=round, >=stealth]
      % Parâmetros
      \def\W{2.6}   % meia-largura das bases
      \def\H{5.0}   % altura dos cones
      \def\grayfill{gray!70}

      % Cone superior (cinza)
      \fill[\grayfill]
        (-\W, \H) -- (\W, \H) -- (0,0) -- cycle;

      % Cone inferior (cinza)
      \fill[\grayfill]
        (-\W,-\H) -- (\W,-\H) -- (0,0) -- cycle;

      % Setas vermelhas (retas a partir do centro)
      \draw[red, ultra thick, ->] (0,0) -- (5, 3);
      \draw[red, ultra thick, ->] (0,0) -- (5,-3.5);
      \draw[red, ultra thick, ->] (0,0) -- (-5, 4);
      \draw[red, ultra thick, ->] (0,0) -- (-5,-2);
    \end{tikzpicture}
    \caption{The red arrows represent the allowed direction for the propagation of the singularity.}
    \label{fig:cones-direcoes}
\end{figure}
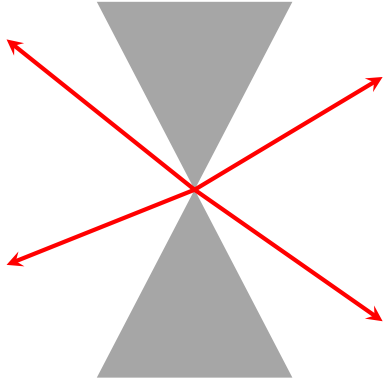

The closure of $V_\pm$ is denoted by $\overline{V}_\pm$. The first property is a natural requirement of symmetry, while the second is a technical condition ensuring the consistency of the theory. A detailed discussion can be found in \cite{book}. Since $f_n$ has compact support, no additional decay conditions on $h\in \mathcal{C}$ are necessary.  

\subsubsection{The set of local fields}

The subset of \textbf{local fields} is defined as finite linear combinations of functionals of the form (\ref{Standard Field}) multiplied by tast functions. Explicitly:
\begin{align}
    \mathcal{F}_{\text{loc}}:=\left\{\sum_{i=1}^K\int dx A_i(x) g_i(x)\,\bigg|\, A_i\in\mathcal{P},\; g_i\in \mathcal{D}(\mathbb{M}),\; K<\infty\right\}.
\end{align}
Most of perturbative quantum field theory is developed within $\mathcal{F}_{\text{loc}}$.  

\subsubsection{Derivative of functionals}

We now introduce the functional derivative $\frac{\delta}{\delta \phi(x)}$. For a field $F\in\mathcal{F}$,
\[
F=f_0+\sum_{n=1}^N\int dx_1\cdots dx_n\, f_n(x_1,\cdots ,x_n)\phi(x_1)\cdots \phi(x_n),
\]
its $k$-th derivative is
\begin{align}
    &\frac{\delta^kF}{\delta \phi(y_1)\cdots \delta\phi(y_k)}\nonumber\\
    &:=\sum_{n=k}^N\frac{n!}{(n-k)!}\int dx_1\cdots  dx_{n-k}\phi(x_1)\cdots \phi(x_{n-k})f_n(y_1,\cdots ,y_k,x_1,\cdots ,x_{n-k}).
\end{align}

The support of $F$ is then defined via its derivative:
\begin{align}
    \supp F:=\supp\frac{\delta F}{\delta\phi(\cdot)}\equiv\overline{\bigcup_{h\in\mathcal{C}}\supp\frac{\delta F}{\delta \phi(\cdot)}(h)}.
\end{align}

This definition is consistent with the standard property
\begin{gather}
    \frac{\delta\phi(x)}{\delta\phi(y)}=\delta(x-y).
\end{gather}

\subsubsection{Classical product}

Both $\mathcal{F}$ and $\mathcal{F}_{\text{loc}}$ are vector spaces. Since no topology is imposed, convergence is understood pointwise: $F_n\to F$ if
\begin{align}
    \lim_{n\rightarrow\infty} F_n(h)=F(h),\quad \forall h\in\mathcal{C}.
\end{align}

The product of fields is defined pointwise as
\begin{align}
    F\cdot G\equiv FG:h\mapsto F(h)G(h).
\end{align}

We also introduce conjugation and parity transformations:
\begin{align}
    F^*:=\sum_n\int dx_1\cdots  dx_n\,\phi(x_1)\cdots \phi(x_n)\overline{f}_n(x_1,\cdots ,x_n),
\end{align}
\begin{align}
    &\alpha(F):=\sum_n\int dx_1...dx_n\alpha(\phi(x_1))...\alpha(\phi(x_n))f_n(x_1,...,x_n),\nonumber\\ &\alpha(\phi(x))=-\phi(x).
\end{align}

\textbf{Technical note.} To show that the classical product is well-defined, one must verify that the defining conditions (\ref{definef}) are preserved. Symmetry follows directly, while the wave front set condition is ensured by Hörmander’s criterion.   

\subsubsection{Classical theory}

We now turn to the free theory. The action is
\begin{align}
    S_0:=\int dx \frac{1}{2}\left(\partial_\mu \phi\partial^\mu\phi-m^2\phi^2\right)\equiv \int dx\,L_0(x).
\end{align}

Although $S_0\notin\mathcal{F}_{\loc}$, this does not present an issue, since the relevant quantity is its derivative,
\[
\frac{\delta S_0}{\delta \phi(x)}=-\left(\partial^2+m^2\right)\phi(x).
\]

\subsubsection{Generalized Lagrangian}

The interacting part of the theory is described by a \textbf{generalized Lagrangian}, defined as a map
\begin{align}
    \mathcal{L}:\mathcal{D}(\mathbb{M})\rightarrow\mathcal{F}_{\text{loc}}
\end{align}
satisfying
\begin{align}
    &\supp\mathcal{L}(f)\subseteq \supp f,\quad \forall f\in\mathcal{D}(\mathcal{M}),\quad  \mathcal{L}(0)=0, \nonumber\\
    &\mathcal{L}(f+g+h)=\mathcal{L}(f+g)-\mathcal{L}(g)+\mathcal{L}(g+h), \quad \text{if } \supp(f)\cap\supp(h)=\emptyset.
\end{align}

This condition ensures additivity when supports do not overlap. Equivalence of Lagrangians is defined by
\begin{align}
    L_1-L_2=c+\partial_\mu A^\mu,
\end{align}
which corresponds to the usual freedom of adding constants or total divergences.  

The interacting action is then written as
\begin{align}
    S=\int dx\mathcal{L}_{\text{int}}(x);\quad \mathcal{L}_{\text{int}}(x)=-\kappa g(x)L_{\text{int}}(x).
\end{align}

The field equation is obtained from
\begin{align}
    \frac{\delta(S+S_0)}{\delta\phi(x)}=0,
\end{align}
which reproduces the Euler–Lagrange equation when $S$ depends only on $\phi$ and $\partial^\mu\phi$.

The solution space is
\begin{align}
    \mathcal{C}_{S+S_0}:=\left\{h\in\mathcal{C}\,\bigg|\,\frac{\delta(S_0+S)}{\delta\phi(x)}(h)=0,\;\forall x\in \mathbb{M}\right\}.
\end{align}

Fields restricted to this space are called \textbf{on-shell}, while unrestricted ones are \textbf{off-shell}.  

\subsubsection{Poisson algebra of the free theory}

We now introduce the Poisson algebra of the free theory. Following the construction of Peierls \cite{Pierls}, the Poisson bracket of $F,G\in\mathcal{F}$ is
\begin{align}
    \{F,G\}:=\int dxdy\, \frac{\delta F}{\delta\phi (x)}\Delta(x-y)\frac{\delta G}{\delta \phi(y)}, \quad \Delta(x-y)=\Delta^{\ret}(x-y)-\Delta^{\ret}(y-x).
    \label{Poisson Bracket}
\end{align}

Here $\Delta^{\ret}(x)$ is the retarded propagator, and $\Delta(x)$ the commutator function:
\begin{align}
    \Delta^{\ret}(x)&:=\frac{1}{(2\pi)^d}\int d^dp\, \frac{e^{-ipx}}{p^2-m^2+ip^00},\nonumber\\
    \Delta(x)&:=\frac{-i}{(2\pi)^{d-1}}\int d^dp\, \sgn(p^0)\delta(p^2-m^2)e^{-ipx}.
\end{align}

The Poisson bracket satisfies the following properties:
\begin{itemize}
    \item[i)] $\{F,G\}\in\mathcal{F}$ due to the wave front set conditions;
    \item[ii)] bilinearity;
    \item[iii)] antisymmetry: $\{F,G\}=-\{G,F\}$;
    \item[iv)] Leibniz rule: $\{F,GH\}=\{F,G\}H+\{F,H\}G$;
    \item[v)] Jacobi identity.
\end{itemize}

This structure generalizes the canonical Poisson brackets. In particular, choosing $F=\phi(x)$ and $G=\partial_{x^0}\phi(x)$ recovers the standard equal-time relations.

%%%%%%%%%%%%%%%%%%%%%%%%%%%%%%%%%%%%%%%%%%%%%%%%%%%%%%%%%%%%%%%%%%%%%
\section{Quantization}

\subsection{Introduction}

The transition from classical to quantum theory is achieved through a quantization procedure. The most common approaches in the physics literature are canonical quantization and path integral quantization. \\

However, canonical quantization is not the most suitable method for the formalism adopted here. Its main limitation, for our purposes, is the requirement to promote fields to operators (in addition to other technical issues that are not central to the present discussion; see, for example, \cite{GROENEWOLD1946405}). In our framework, the Fock space is introduced only at a later stage, so the promotion of fields to operators is not desirable — we want fields to remain fields. The solution is provided by \textit{deformation quantization}. This method has its origin in the work of von Neumann \cite{VonNeumann1932}, but was formally developed only in the late 1970s \cite{BAYEN1978111}. For pedagogical reviews, see \cite{Hirshfeld_2002,Hancock_2004}; for further developments, consult \cite{dito2002deformationquantizationgenesisdevelopments,gutt2000variationsdeformationquantization,ZACHOS_2002}.

\subsection{Deformation quantization}

Just as the Peierls bracket can be seen as a generalization of the classical Poisson bracket, deformation quantization provides a generalized notion of commutators. The idea is to expand commutators (e.g., $[q,p]=i\hbar$) in a power series in $\hbar$. To implement this, we introduce the \textit{star product} $\star_\hbar:\mathcal{F}\times\mathcal{F}\rightarrow\mathcal{F}\llbracket\hbar\rrbracket$, defined by the following conditions:

\begin{itemize}
    \item[a)] bilinear in its arguments, \\
    \item[b)] associative, \\
    \item[c)] $F\star_\hbar G \to F\cdot G$ as $\hbar\to 0$, \\
    \item[d)] $\dfrac{F\star_\hbar G - G\star_\hbar F}{i\hbar} \equiv \dfrac{[F,G]_{\star_\hbar}}{i\hbar} \to \{F,G\}$ as $\hbar \to 0$. 
\end{itemize}

Although the discussion at this stage is abstract, deformation quantization naturally appears already in non-relativistic quantum mechanics. For an accessible introduction, see \cite{Hirshfeld_2002,Hancock_2004}. The main advantage of this approach is that it allows the quantization of systems with complicated classical structures, which is particularly relevant in curved spacetimes. Since we restrict attention to Minkowski spacetime and polynomial fields, the star product can be defined with stronger axioms as a formal power series in $\hbar$. Explicitly, for $F,G\in\mathcal{F}$ we define:

\begin{align}
    &F\star_\hbar G:=\sum_{n=0}^\infty \frac{\hbar^n}{n!}\int dx_1\cdots dx_ndy_1\cdots dy_n\nonumber\\
    &\times \frac{\delta ^nF}{\delta\phi(x_1)\cdots \delta\phi(x_n)}\prod_{l=1}^nH_m(x_l-y_l)\frac{\delta^nG}{\delta\phi(y_1)\cdots\delta\phi(y_n)}.
\end{align}

The star product satisfies the following properties:

\begin{enumerate}[label=\alph*)]
    \item Bilinearity,  
    \item Associativity,  
    \item $F \star_\hbar G \to F \cdot G$ as $\hbar \to 0$,  
    \item $\dfrac{1}{i\hbar}[F,G]_\star \to \{F,G\}$ as $\hbar \to 0$,  
    \item $(\partial^2 + m^2)H_m(x-y) = 0 \iff (\partial^2_x + m^2)(\phi(x) \star_\hbar \phi(y))_0 = 0$ for $\phi(x) \in F_\mathcal{C_{S0}}$,  
    \item Lorentz invariance: $H_m(\Lambda z) = H_m(z)$ for all $\Lambda \in \mathcal{L}_+^\uparrow$,  
    \item Existence of all powers $H^l_m(y-x)$,  
    \item $\overline{H}_m(x) = H_m(-x)$, equivalently $(F\star_\hbar G)^* = G^*\star_\hbar F^*$.  
\end{enumerate}

\textbf{Remark.} For polynomial fields, the series above truncates after finitely many terms. For non-polynomial fields, additional care is needed to ensure convergence, but a consistent definition of the star product remains possible.

\subsubsection{Wightman two-point function}

In the framework of perturbative quantum field theory in Minkowski spacetime, the appropriate choice of $H_m$ is the Wightman two-point function:

\begin{align}
    H_m(z)&\equiv \Delta^+_m(z):=\frac{1}{(2\pi)^{d-1}}\int d^dp\,\theta(p^0)\delta(p^2-m^2)e^{-ipz}\nonumber\\
    &=\frac{1}{(2\pi)^{d-1}}\int d^{d-1}\vec{p}\,\frac{e^{-i\omega_pz^0+i\vec{p}\,\vec{z}}}{2\omega_p},\quad \omega_p:=\sqrt{(\vec{p})^2+m^2}.
\end{align}

Observe that

\begin{align}
    \Delta^+_m(z)-\Delta^+_m(-z)\equiv i\Delta_m(z)=i(\Delta^{\ret}(z)-\Delta^{\ret}(-z)).
\end{align}

From now on, we omit the explicit $\hbar$ in the notation $\star_\hbar$ for readability. A detailed proof that the star product defined above exists and satisfies the axioms can be found in Chapter 2 of \cite{book}. Importantly, for $F=\phi(x)$ and $G=\partial_{x^0}\phi(x)$, the construction reproduces the canonical commutation relations. \\

\subsection{States}

(The text in this subsection follows closely Section 2.5 of \cite{book}.)  

We denote by $\mathcal{F}_\hbar$ the space of fields that are polynomial in $\hbar$:

\begin{align}
    \mathcal{F}_\hbar:=\left\{\sum_{s=0}^SF_s\hbar^s\;\big|\; F_s\in\mathcal{F},\; S<\infty\right\}.
\end{align}

A state $\omega$ on the algebra $\mathcal{A}\equiv \mathcal{A}_\hbar:=(\mathcal{F}_\hbar,\star)$ is defined as

\begin{align}
    \omega:\begin{cases}
        \mathcal{A}\rightarrow\mathbb{C}\\
        F\mapsto \omega(F)\equiv \omega(F)_\hbar
    \end{cases}
\end{align}

where $\omega$ may itself be polynomial in $\hbar$, i.e. $\omega=\sum_{n=0}^{N}\omega_n\hbar^n$ for some finite $N$. It satisfies:

\begin{itemize}
        \item Linearity: $\omega(F + \alpha G) = \omega(F) + \alpha \omega(G)$, $\forall F, G \in \mathcal{A}$, $\alpha \in \mathbb{C}$,
        \item Reality: $\omega(F^*)_\hbar = \overline{\omega(F)}_\hbar$, $\forall F \in \mathcal{A}$, $\forall \hbar > 0$,
        \item Positivity: $\omega(F^* \star F)_\hbar \geq 0$, $\forall F \in \mathcal{A}$, $\forall \hbar > 0$,
        \item Normalization: $\omega(1) = 1$, with $1\in\mathcal{F}_\hbar$ the constant functional $1(h)=1$ for all $h\in\mathcal{C}$.
\end{itemize}

Consequences:

\begin{enumerate}[label=\alph*)]
    \item If $F^*=F$, then $\omega(F)\in\mathbb{R}$,
    \item Linearity implies $\omega(F) = \sum_{r,s} \omega_s(F_r) \hbar^{r+s}$ for $F=\sum_r F_r \hbar^r$, and the sum is finite.
\end{enumerate}

In particular, the vacuum state $\omega_0$ and coherent states can be defined. Given $F=f_0+\sum_{n\geq1}\langle f_n,\phi^{\otimes n}\rangle$, we set

\begin{align}
    \omega_0(F):=f_0.
\end{align}

\subsection{Bijection of on-shell quantized fields and normal ordered products}\label{Bijection of on-shell quantized fields and normal order products}

We now recall a key structural result (see p.65 of \cite{book}):  

There is an algebra isomorphism $\Phi(F)$ given by

\begin{align}
    F&=\sum_{n=0}^N \int dX_n \phi(x_1)...\phi(x_n)f_n(x_1,...,x_n)\\
    &\mapsto \Phi(F)=\sum_{n=0}^N \int dX_n :\phi^{\op}(x_1)...\phi^{\op}(x_n):f_n(x_1,...,x_n),
\end{align}
where $\phi^{\op}(x)$ represents the field operator in the usual Fock space and $:\bullet:$ is the normal ordering of the field. The map $\Phi$ respects

\begin{align}
    \Phi(F)\cdot\Phi(G)=\Phi(F\star G).
\end{align}

Using the map $\Phi$ we can construct a dictionary between the usual quantum field theory done in the Fock space and the quantum field theory in the space of fields. The best way to see it is using examples.

\subsection{Examples}

\subsubsection{Vacuum state}

In the definition of the star product, the vacuum state was specified as
\[
\omega_0\!\left(f+\sum_{n=0}^N \int dX_n f(x_1,\dots,x_n)\phi(x_1)\dots\phi(x_n)\right)=f.
\]
This coincides with the Fock vacuum, since
\begin{align}
    \langle\Omega|f+\sum_{n=0}^N \int dX_n f(x_1,\dots,x_n):\phi^{\op}(x_1)\dots\phi^{\op}(x_n):\Omega\rangle = f.
\end{align}

\subsubsection{Scattering in the free theory}

\paragraph{Two-particle ``scattering'' in the free theory.}

Consider the amplitude for one incoming and one outgoing particle:
\begin{align}
    \langle\Omega a^*(\vec{p})|a^*(\vec{q})\Omega\rangle,
\end{align}
where $a^*(p),a(q)$ are the standard creation and annihilation operators on Fock space, satisfying
\begin{gather}
    [a^*(\vec{p}),a(\vec{q})]=2\hbar\omega_p\delta(\vec{p}-\vec{q}),\quad [a(\vec{p}),a(\vec{q})]=[a^*(\vec{p}),a^*(\vec{q})]=0.
\end{gather}

Moving $a^*(p)$ across, we find
\begin{align}
    \langle\Omega a^*(p)|a^*(q)\Omega\rangle=\langle\Omega|a(p) a^*(q)\Omega\rangle,
\end{align}
and thus
\begin{align}
    a(p)a^*(q) &= a^*(q)a(p)+[a(p),a^*(q)] \\
               &= a^*(q)a(p)+2\hbar\omega_p\delta(\vec{p}-\vec{q}).
\end{align}
Hence,
\begin{align}
    \langle\Omega a^*(p)|a^*(q)\Omega\rangle=0+2\hbar\omega_p\delta(\vec{p}-\vec{q}).
\end{align}

To compare with the functional formalism, we express $a^*(p)\Omega$ in terms of fields in configuration space. Introducing the operator-valued field
\begin{gather}
    \phi^{\op}(x):=\frac{1}{(2\pi)^{\frac{d-1}{2}}}\int\frac{d\vec{p}}{2\omega_p}\Big(e^{ipx}a^*(\vec{p})+e^{-ipx}a(\vec{p})\Big)\Big\rvert_{p^0=\omega_p},
\end{gather}
we compute
\begin{align}
    \phi^{\op}(x)\Omega
    &=\frac{1}{(2\pi)^{\frac{d-1}{2}}}\int \frac{d\vec{p}}{2\omega_p}e^{ipx} a^*(\vec{p})\Omega.
\end{align}

Applying the inverse Fourier transform yields
\begin{align}
    a^*(\vec{p})\Omega
    =\frac{2\omega_p}{(2\pi)^{\frac{d-1}{2}}}\int d\vec{x}\,e^{-ipx}\phi^{\op}(x)\Omega\bigg\rvert_{p^0=\omega_p}.
\end{align}

Finally, the amplitude becomes
\begin{align}
    \mathcal{T}_2=\langle \Omega a^*(\vec{p})|a^*(\vec{q})\Omega\rangle
    =\frac{4\omega_p\omega_q}{(2\pi)^{d-1}}\int d\vec{x}d\vec{y}\, \langle\Omega|e^{ipx-iqy}:\phi^{\op}(\vec{x})::\phi^{\op}(\vec{y}):\Omega\rangle.
\end{align}

In the functional formalism this is expressed as
\begin{align}
    \mathcal{T}(p,q)=\frac{4\omega_p\omega_q}{(2\pi)^{d-1}}\int d\vec{x}d\vec{y}\,\omega_0(\phi(x)\star\phi(y))e^{ipx-iqy}\bigg\rvert_{p^0=\omega_p,q^0=\omega_q}.
\end{align}

Since
\begin{align}
    \phi(x)\star\phi(y)&=\phi(x)\phi(y)+\hbar\Delta^+(x-y), \\
    \omega_0(\phi(x)\star\phi(y))&=\hbar\Delta^+(x-y)=\frac{\hbar}{(2\pi)^{d-1}}\int d\mu_k e^{-ik(x-y)}\Big\rvert_{k_0=\omega_k},
\end{align}
evaluating the integrals gives
\begin{align}
    \mathcal{T}(p,q)=2\omega_p\hbar\delta(\vec{p}-\vec{q}).
\end{align}

This result corresponds to the free Feynman diagram:

\begin{figure}[H]
\centering
\begin{tikzpicture}
  \begin{feynman}
    \vertex (p) {\(p\)};
    \vertex [right=3cm of p] (q) {\(q\)};
    \diagram* {
      (p) -- [scalar] (q),
    };
  \end{feynman}
\end{tikzpicture}
\end{figure}

\subsection*{General structure of scattering amplitudes}

The previous computations illustrate the equivalence of both formalisms in the free case. More generally, scattering amplitudes take the form
\begin{align}
    \langle \Omega a^*(\vec{p}_1)\dots a^*(\vec{p}_n)|Sa^*(\vec{q}_1)\dots a^*(\vec{q}_m)\Omega\rangle.
\end{align}

Translating into the functional picture requires Fourier inversion. Each incoming particle carries a factor
\begin{align}
    \frac{2\omega_{p_i}}{(2\pi)^{\frac{d-1}{2}}}\int d\vec{x}_ie^{ip_ix_i}\bigg\rvert_{p^0_i=\omega_{p_i}},
\end{align}
and each outgoing particle contributes
\begin{align}
    \frac{2\omega_{q_i}}{(2\pi)^{\frac{d-1}{2}}}\int d\vec{y}_ie^{-iq_iy_i}\bigg\rvert_{q^0_i=\omega_{q_i}}.
\end{align}

The expressions represent the Feynman rules for scattering of scalar particles. Feynman rules for other types of fields can be found at \cite{grande2025}.

%%%%%%%%%%%%%%%%%%%%%%%%%%%%%%%%%%%%%%%%%%%%%%%%%%%%%%%%%%%%%%%%%%%%%
\section{T-product}

\subsection{Introduction}

The next step toward a complete scattering theory is the construction of the scattering matrix:
\begin{align}
    S(F):=1+\sum_{n=1}^\infty\frac{i^n}{n!\hbar^n}T_n\left(F^{\otimes n}\right).
\end{align}

Following the inductive and axiomatic approach of Epstein–Glaser \cite{epstein1973role}, the operators $T_n$ are constructed recursively. Our presentation is guided primarily by \cite{book}, Chapter 3.3.

We begin by introducing the axioms. Since the scope of this article is expository, we restrict ourselves to the axioms that will be explicitly applied in the forthcoming discussion (basic axioms). The second group of axioms (renormalization conditions), although essential for a consistent formulation, will only be outlined.

\subsubsection{Basic axioms}

The basic axioms can be summarized as follows:  \\
(I) Linearity,  \\
(II) Symmetry in the arguments,  \\
(III) Initial condition $T_1(F)=F$,  \\
(IV) Causality:
\begin{align}
    &T_n(A_1(x_1),...,A_n(x_n))\nonumber\\
    =&T_k(A_{\pi(1)}(x_{\pi(1)}),....,A_{\pi(k)}(x_{\pi(k)}))\star T_{n-k}(A_{\pi(k+1)}(x_{\pi(k+1)}),...,A_{\pi(n)}(x_{\pi(n)}))
    \label{Causal Fatorization T}
\end{align}
whenever $\{x_{\pi(1)},...,x_{\pi(k)}\}\cap (\{x_{\pi(k)},...,x_{\pi(n)}\}+\overline{V}_-)=\emptyset$.

\subsubsection{Renormalization conditions}

The renormalization conditions are listed below:

\begin{itemize}
    \item[(v)] \textbf{Field independence}
    \item[(vi)] \textbf{$\ast$-structure (unitarity of $S$) and field parity}
    \item[(vii)] \textbf{Poincaré covariance}
    \item[(viii)] \textbf{Off-shell field equation}
    \item[(ix)] \textbf{Sm-expansion}
    \item[(x)] \textbf{$\hbar$-dependence}
\end{itemize}

Further details can be found in Chapter 3 of \cite{book}.

\subsection{Construction of $T$ product}

We first construct $T_2(L(x_1),L(x_2))$ and then briefly discuss the generalization to higher orders. The construction relies on dividing the domain of $T_2$ such that the causality axiom reduces the problem to the product of $T_1(L(x_1))=L(x_1)$ and $T_1(L(x_2))=L(x_2)$. The space $\mathbb{R}^2$ can be partitioned into the following regions:

\begin{itemize}
    \item[a)] $x_1\notin (x_2+\overline{V}_-)$
    \item[b)] $x_2\notin (x_1+\overline{V}_-)$
    \item[c)] $x_1=x_2$
\end{itemize}

In region (a), the causality axiom yields:
\begin{align}
    T_2(L(x_1),L(x_2))=T_1(L(x_1))\star T_1(L(x_2))=L(x_1)\star L(x_2).
\end{align}

In region (b), the same formula holds with indices exchanged:
\begin{align}
    T_2(L(x_1),L(x_2))=L(x_2)\star L(x_1).
\end{align}

For $x_1\neq x_2$, the result can be summarized using the step functions $\theta(x^0)$:
\begin{align}
    T_2(L(x_1),L(x_2))&=L(x_1)\star L(x_2)\theta(x^0_2-x_1^0)+L(x_2)\star L(x_1)\theta(x_1^0-x_2^0)\nonumber\\
    &\equiv L(x_1)\star_F L(x_2).
    \label{T_2}
\end{align}

Here, $\star_F$ denotes the star product with $\Delta^+$ replaced by the Feynman propagator $\Delta^F$, defined by
\begin{gather}
    \Delta^F(x)=\Delta^+(x)\theta(x^0)+\Delta^+(-x)\theta(-x^0).
\end{gather}

The remaining case (c) requires extending the distribution in (\ref{T_2}), originally defined on $\mathbb{R}^2\setminus\Delta_2$, to the whole space $\mathbb{R}^2$. The procedure of extending the domain of the distribution is what is usually called ``renormalization'' in quantum field theory. We stress that the procedure presented here is not plagued by divergent integrals, rather non-uniqueness of the extension of the distribution. 

The extension of a distribution $t^0\in\mathcal{D}'(\mathbb{R}^d\setminus\{0\})$ is characterize by a number that measures ``the strength of divergence of a distribution'' called \textit{scalling degree} and defined through 
\begin{align}
    \sd(t^0):=\inf_{r\in\mathbb{R}}\,\{ \lim_{r\rightarrow0}\varepsilon^rt(\varepsilon r)=0\}.
\end{align}

The extension of the distribution is related to the scaling degree by the following theorem:\\

Let $t^0\in\mathcal{D}'(\mathbb{R}^d\setminus\{0\})$ and $t\in\mathcal{D}'(\mathbb{R}^d)$ be an extension of $t^0$. If $\sd(t^0)<d$, then $t$ is uniquely determined by $t^0$. If $\sd(t^0)\geq d$, the extension is not unique, and any two extensions differ by
\begin{align}
    t=t'+\sum_{|\alpha|\leq\sd(t^0)-d} C_\alpha\partial^\alpha\delta(x),
\end{align}
where $C_\alpha\in\mathbb{C}$, $\alpha$ is a multi-index. Basically, the theorem connects the divergent integrals of the usual quantum field theory with non-unique distributions. The more ``divergent'' the integral is more freedom we have to define its extension. A pedagogical example is provided in the appendix.

A proof can be found in \cite{KlausQFTNotes}, Theorem V.4.  \\

In particular, in $d=4$ dimensions the Feynman propagator and the Wightman two-point function satisfy $\sd(\Delta^F)=\sd(\Delta^+)=2$. Thus, powers of $\Delta^F$ or $\Delta^+$ generally require non-trivial renormalization. This explains the emergence of ``infinities'' in quantum field theory: integrals are not divergent in the distributional framework, but rather ambiguously defined. 

\subsubsection{Construction of the unrenormalized $T$-product on $\mathbb{M}^n\setminus\Delta_n$}

The unrenormalized $T$-product is constructed inductively. Assume that $T_m(L^{\otimes m})$ has been defined for all $m\in{1,\dots,n-1}$. To define $T_n(L^{\otimes n})$, one partitions the domain into subsets
\begin{align}
C_I:=&{(x_1,...,x_n)\in\mathbb{M}^n ,|, x_i\notin(x_j+\overline{V}_-), \forall i \in I, j\in I^C},\nonumber\\
&\text{with } I\subset{1,...,n}, , I\neq {1,...,n}, , I\neq\emptyset.
\end{align}

Each point $(x_1,...,x_n)\in\mathbb{M}^n\setminus\Delta_n$ belongs to at least one such set, yielding
\begin{align}
\bigcup_I C_I=\mathbb{M}^n\setminus\Delta_n.
\end{align}

For a given $I$, causality implies
\begin{align}
&T_n^0(A_1(x_1),...,A_n(x_n))\nonumber\\
=&T_{|I|}(A_{i_1}(x_{i_1}),...,A_{i_{|I|}}(x_{i_{|I|}}))\star T_{|I^c|}(A_{j_1}(x_{j_1}),...,A_{j_{|I^c|}}(x_{j_{|I^c|}})).
\end{align}

The extension to the thin diagonal is carried out via the Epstein–Glaser procedure. Detailed treatments can be found in \cite{epstein1973role}, in the book by Dütsch \cite{book}, and in \cite{Brunetti_2000}. For the purposes of this work, the formula above suffices, as the extension to the diagonal is straightforward and applicable to the entire domain.

\subsection{Tree-level amplitude for $L_{int}=-(\frac{\lambda}{4!}) \int dx, g(x)\phi^4(x)$}

We now employ the formalism introduced above to compute the $S$-matrix of the $\lambda\phi^4$ theory. As emphasized previously, the interaction must be multiplied by a test function $g$ to ensure mathematical well-definedness. For the scattering amplitude, however, the adiabatic limit $g(x)=1 ;\forall x$ will be shown to exist. Consider
\begin{align}
S(L_{int})\equiv T\Big(e^{\frac{iL_{int}}{\hbar}}\Big)=\frac{i}{\hbar}T_{1}(L_{int})+\frac{i^2}{2!\hbar^2}T_2(L_{int},L_{int})+\mathcal{O}\Big(\big(\frac{\lambda}{4!}\big)^3\Big).
\end{align}

At first order, one obtains
\begin{align}
\frac{i}{\hbar}T_1\Big(-\frac{\lambda}{4!}\int dx_1\,g(x_1)\phi^4(x_1)\Big)=\frac{-i(\frac{\lambda}{4!})}{\hbar}\int dx_1\,g(x_1)\phi^4(x_1).
\end{align}

Using the previous results, the scattering amplitude $\mathcal{T}$ can be computed. In the Fock space formalism, fields in the $T$-product are normal ordered \cite{Scharf:FiniteQED}, allowing the application of the theorem in Section (\ref{Bijection of on-shell quantized fields and normal order products}). For $2 \to 2$ scattering, the first-order contribution in $\lambda$ reads
\begin{align}
\mathcal{T}1&=\left(\prod_{i=1}^4\frac{2\omega_{p_i}}{(2\pi)^{\frac{d-1}{2}}}\right)\int d\vec{Y}_4dx\nonumber\\
&\times\omega_0\Big(e^{ip_1y_1+ip_2y_2}\phi(y_1)\phi(y_2)\star \frac{-i\lambda}{4!\hbar}g(x)\phi^4(x)\star e^{-ip_3y_3-ip_4y_4}\phi(y_3)\phi(y_4)\Big).
\end{align}

The computation proceeds by first evaluating the star product. For instance:
\begin{align}
\phi(y_1)\phi(y_2)\star\phi^4(x)&=\phi(y_1)\phi(y_2)\phi^4(x)\nonumber\\
&+4\hbar \left(\phi(y_1)\Delta^+(y_2-x)\phi^3(x)+\phi(y_2)\Delta^+(y_1-x)\phi^3(x)\right)\nonumber\\
&+\frac{4\cdot3\hbar^2}{2!}\bigg(\Delta^+(y_1-x)\Delta^+(y_2-x)\phi^2(x)\nonumber\\
&+\Delta^+(y_2-x)\Delta^+(y_1-x)\phi^2(x)\bigg).
\end{align}

Since only vacuum expectation values are needed, only terms proportional to $\phi^2$ contribute when evaluating $(\phi(y_1)\phi(y_2)\star\phi^4(x))\star\phi(y_3)\phi(y_4)$, giving
\begin{align}
\phi^2(x)\star\phi(y_3)\phi(y_4)&=\phi^2(x)\phi(y_3)\phi(y_4)\nonumber\\
&+2\hbar \big(\phi(x)\Delta^+(x-y_3)\phi(y_4)+\phi(x)\Delta^+(x-y_4)\phi(y_3)\big)\nonumber\\
&=2\hbar^2\Delta^+(x-y_4)\Delta^+(x-y_3).
\end{align}

Thus,
\begin{align}
&\omega_0(\phi(y_1)\phi(y_2)\star\phi^4(x)\star\phi(y_3)\phi(y_4))\nonumber\\
&=24\hbar^4\Delta^+(y_1-x)\Delta^+(y_2-x)\Delta^+(x-y_3)\Delta^+(x-y_4),
\end{align}
and
\begin{align}
&\mathcal{T}1=-i\lambda\hbar^3\left(\prod_{i=1}^4\frac{2\omega_{p_i}}{(2\pi)^{\frac{d-1}{2}}}\right)e^{ip_1y_1+ip_2y_2-ip_3y_3-ip_4y_4}\nonumber\\
&\times \int d\vec{Y}_4dx\,g(x) \Delta^+(y_1-x)\Delta^+(y_2-x)\Delta^+(x-y_3)\Delta^+(x-y_4).
\end{align}

The integrals over $\vec{Y}_4$ can be evaluated explicitly. Integration over $x$ requires attention due to the test function $g$, and the adiabatic limit $g(x)\rightarrow 1;\forall x$ must be justified. Relevant integrals include
\begin{align}
&\int d\vec{y}_1\,\Delta^+(y_1-x)e^{ip_1y_1}=\frac{e^{ip_1x}}{2\omega{p_1}},\
&\int d\vec{y}_3\,\Delta^+(x-y_3)e^{-ip_3y_3}=\frac{e^{-ip_3x}}{2\omega{p_3}}.
\end{align}

Collecting all contributions, one finds
\begin{align}
\mathcal{T}1&=-i\lambda\hbar^3\left(\prod_{i=1}^4\frac{2\omega_{p_i}}{(2\pi)^{\frac{d-1}{2}}}\right)\left(\prod_{i=1}^4\frac{1}{2\omega_{p_i}}\right)\int dx\, g(x)e^{i(p_1+p_2-p_3-p_4)x}\nonumber\\
&\overset{g\rightarrow1}{=}-i\frac{\lambda}{(2\pi)^{d-2}}\hbar^3\delta(p_1+p_2-p_3-p_4),
\end{align}
which reproduces the Feynman rule for the $\lambda\phi^4$ vertex in momentum space.

\subsection{Second-order corrections}

The second-order contribution to the scattering amplitude is obtained from
\begin{align}
\frac{i^2}{2\hbar^2}T_2(L_{int},L_{int})=-\frac{1}{2\hbar^2}L_{int}\star_F&L_{int}\nonumber\\
=-\frac{\lambda^2}{2(4!)^2\hbar^2}\int& dxy\,g(x)g(y) \phi^4(x)\star_F\phi^4(y)\nonumber\\
=-\frac{\lambda^2}{2(4!)^2\hbar^2}\int& dxy\,g(x)g(y)\bigg(\phi^4(x)\phi^4(y)\nonumber\\
+&16\hbar\phi^3(x)\Delta^F(x-y)\phi^3(y)\nonumber\\
+&\frac{\hbar^2}{2}144\phi^2(x)(\Delta^F(x-y))^2\phi^2(y)\nonumber\\
+&\frac{\hbar^3}{6}576\phi(x)(\Delta^F(x-y))^3\phi(y)\nonumber\\
+&\frac{\hbar^4}{24}576(\Delta^F(x-y))^4\bigg).
\end{align}

Only terms with four powers of $\phi$ contribute to the amplitude:
\begin{align}
\mathcal{T}2=
\omega_0\Big(\overline{\phi(p_1)\phi(p_2)}&\star \Big(\frac{-36\lambda^2\hbar^2}{(4!\hbar)^2}\int dxdy\,g(x)g(y)\phi^2(x)\phi^2(y)(\Delta^F(x-y))^2\Big)\nonumber\\
&\star\phi(p_3)\phi(p_4)\Big),
\end{align}
where
\begin{align}
&\overline{\phi(p_1)\phi(p_2)}\equiv\frac{4\omega{p_1}\omega_{p_2}}{(2\pi)^{d-1}}\int d\vec{x}_1d\vec{x}_2\, e^{ip_1x_1+ip_2x_2}\phi(x_1)\phi(x_2),\nonumber\\
&\phi(p_3)\phi(p_4)\equiv \frac{4\omega{p_3}\omega{p_4}}{(2\pi)^{d-1}}\int d\vec{x}_3d\vec{x}_4\, e^{-ip_3x_3-ip_4x_4}\phi(x_3)\phi(x_4).
\end{align}

After evaluation of the star product:
\begin{align}
\mathcal{T}_2=&-\frac{144\lambda^2\hbar^2}{(2\pi)^{2(d-1)}(4!)^2}\int dxdy\, g(x)g(y)e^{i(p_1+p_2)x-i(p_3+p_4)y} (\Delta^F(x-y))^2\nonumber\\
&-\frac{144\lambda^2\hbar^2}{(2\pi)^{2(d-1)}(4!)^2}\int dxdy\, g(x)g(y)e^{i(p_1-p_3)x+i(p_2-p_4)y} (\Delta^F(x-y))^2\nonumber\\
&-\frac{144\lambda^2\hbar^2}{(2\pi)^{2(d-1)}(4!)^2}\int dxdy\, g(x)g(y)e^{i(p_1-p_4)x+i(p_2-p_3)y} (\Delta^F(x-y))^2,
\end{align}
corresponding to the $s$, $t$, and $u$ channel diagrams

\begin{figure}[h]
\centering
\begin{tikzpicture}[scale=0.8]
  \begin{feynman}
    % Primeiro (canal s)
    \vertex (x1) at (-4,2) {\(x_1\)};
    \vertex (x2) at (-4,-2) {\(x_2\)};
    \vertex (a) at (-2.5,0);
    \vertex (b) at (-1.5,0);
    \vertex (x3) at (0,2) {\(x_3\)};
    \vertex (x4) at (0,-2) {\(x_4\)};
    
    \diagram* {
      (x1) -- [scalar] (a),
      (x2) -- [scalar] (a) -- [scalar, half left, looseness=1.2] (b) -- [scalar] (x4),
      (a) -- [scalar, half right, looseness=1.2] (b),
      (b) -- [scalar] (x3)
    };
    \node at (-2.5,0.7) {\(x\)};
    \node at (-1.5,0.7) {\(y\)};
    
    % Sinal de +
    \node at (1,0) {\Huge +};

    % Segundo (canal t)
    \vertex (y1) at (2,2) {\(x_1\)};
    \vertex (y2) at (2,-2) {\(x_2\)};
    \vertex (c) at (3,1);
    \vertex (d) at (3,-1);
    \vertex (y3) at (4,2) {\(x_3\)};
    \vertex (y4) at (4,-2) {\(x_4\)};

    \diagram* {
      (y1) -- [scalar] (c) -- [scalar] (y3),
      (y2) -- [scalar] (d) -- [scalar] (y4),
      (c) -- [scalar, half left, looseness=1] (d),
      (c) -- [scalar, half right, looseness=1] (d),
    };
    \node at (3.1,1.4) {\(x\)};
    \node at (3.1,-1.4) {\(y\)};

    % Sinal de +
    \node at (5,0) {\Huge +};

    % Terceiro (canal u)
    \vertex (z1) at (6,2) {\(x_1\)};
    \vertex (z2) at (6,-2) {\(x_2\)};
    \vertex (e) at (7,1);
    \vertex (f) at (7,-1);
    \vertex (z3) at (10,2) {\(x_3\)};
    \vertex (z4) at (10,-2) {\(x_4\)};
    
    \diagram* {
      (z1) -- [scalar] (e) -- [scalar] (z4),
      (z2) -- [scalar] (f) -- [scalar] (z3),
      (e) -- [scalar, half left, looseness=0.5] (f),
      (e) -- [scalar, half right, looseness=0.5] (f),
    };
    \node at (7.2,1.4) {\(x\)};
    \node at (7.2,-1.4) {\(y\)};
  \end{feynman}
\end{tikzpicture}
%\caption*{Conjunto 2 – Três diagramas de segunda ordem (\(s\), \(t\), \(u\))}
\end{figure}

In this case, $\sd((\Delta^F)^2)=4$, requiring a non-trivial renormalization. The extension of the distribution is performed using $W$-expansions \cite{book}. The renormalized Feynman propagator is
\begin{align}
\widetilde{(\Delta^F)^2}=\frac{1}{(2\pi)^4}\int dp \frac{2p(p_1+p_2)-(p_1+p_2)^2}{(p^2-m^2+i0)^2((p-p_1-p_2)^2-m^2+i0)}+C,
\end{align}
where $C$ is the counter-term. This expression coincides with standard results in QFT \cite{Cheng:1984vwu}, and $C$ must be constrained to satisfy the axioms of the theory \cite{Pinter_2001}. The complete calculation lies beyond the scope of this article. The renormalization was performed in \cite{Pinter_2001,book,grande2025}.

%%%%%%%%%%%%%%%%%%%%%%%%%%%%%%%%%%%%%%%%%%%%%%%%%%%%%%%%%%%%%%%%%%%%%
\section*{Acknowledgments}
We appreciate the financial support from CAPES during the Master's project \cite{grande2025} from which this work originates. \\

%%%%%%%%%%%%%%%%%%%%%%%%%%%%%%%%%%%%%%%%%%%%%%%%%%%%%%%%%%%%%%%%%%%%%
\appendix

\section{Divergent integrals and the principal value: an example.}

In the main text we argued that divergences do not appear in quantum field theory when one adopts an appropriate mathematical framework. The purpose of this appendix is to provide an explicit example of this statement. Besides its illustrative value, the example is also relevant for practical computations in quantum field theory, as it corresponds to a case of differential renormalization, a method employed in the evaluation of renormalized quantities in QFT \cite{book}, Chapter 3.5. \\

Throughout this example we work in $\mathbb{R}^2$ with smooth functions that decay sufficiently fast at infinity together with all their derivatives. This space is called the Schwartz space and is denoted by $\mathcal{J}(\mathbb{R}^2)$\footnote{``Fast decay'' means that $\lim_{|x|\rightarrow\infty} \|x\|^n \partial^\alpha f(x)=0$ for all $n\in\mathbb{N}$ and for any multi-index $\alpha$.}. This space is larger than the usual space of test functions, but this enlargement does not affect the analysis in the present case. \\

Let $r:=\sqrt{x^2+y^2}$, with $e^{-r^2}\in\mathcal{J}(\mathbb{R}^2)$ and $\frac{1}{r},\frac{1}{r^2}\in\mathcal{J}'(\mathbb{R}^2)$. We can compute

\begin{align}
    \langle\frac{1}{r},e^{-r^2}\rangle=\int dxdy\, \frac{e^{-r^2}}{r}=2\pi \int_0^\infty rdr\, \frac{e^{-r^2}}{r}=\pi^{\frac{3}{2}}.
\end{align}

On the other hand,

\begin{align}
    \langle\frac{1}{r^2},e^{-r^2}\rangle=2\pi \int_0^\infty rdr\, \frac{e^{-r^2}}{r^2}\text{ ``}&=\text{'' }\infty.
\end{align}

The issue in the calculation above is the assumption that the integral must be taken literally. Instead, we may proceed as follows:

\begin{align}
    \frac{d}{dr}\left(\frac{-1}{r}\right)=\frac{1}{r^2}.
\end{align}
Hence, 

\begin{align}
    \langle\frac{1}{r^2},e^{-r^2}\rangle=2\pi \int_0^\infty dr\, \left(\frac{d}{dr}\frac{-1}{r}\right)e^{-r^2}\nonumber\\
    =2\pi\int_0^\infty dr\, \frac{1}{r}\left(\frac{d}{dr}e^{-r^2}\right)=-2\pi^{\frac{3}{2}}.
\end{align}

This calculation exemplifies the notion of the ``principal value'' distribution \cite{hörmander1983analysis}, p.~73. \\

Now, suppose we have constructed the distributions $\frac{1}{r},\frac{1}{r^2}\in\mathcal{J}'(\mathbb{R}^2\setminus\{0\})$ and we are looking for their extensions. First, we need to compute the scaling degree:

\begin{align}
    \sd\left(\frac{1}{r}\right)=1<2\quad \sd\left(\frac{1}{r^2}\right)=2.
\end{align}

Hence, the extensions take the form

\begin{align}
    \widetilde{\frac{1}{r}}=\frac{1}{r};\quad
    \widetilde{\frac{1}{r^2}}=\frac{1}{r}\frac{d}{dr}+C\delta(r).
\end{align}

where $C$ is a counterterm. In the formula for the extension of $\frac{1}{r^2}$ we exploit the trick of rewriting it as a derivative, which allows us to define it as a distribution that can be evaluated as an integral over $\mathcal{J}(\mathbb{R}^2)$. The conclusion is that integrals which ``diverge'' can be corrected by the addition of counterterms, exactly as in standard QFT. The stronger the divergence, the greater the freedom in choosing these counterterms.

%%%%%%%%%%%%%%%%%%%%%%%%%%%%%%%%%%%%%%%%%%%%%%%%%%%%%%%%%%%%%%%%%%%%%

\printbibliography[heading=bibintoc, title={Bibliography}]

\end{document}